\documentclass[manuscript]{acmart}
\AtBeginDocument{%
  }

\setcopyright{acmlicensed}
\copyrightyear{2025}
\acmYear{2025}
\acmDOI{XXXXXXX.XXXXXXX}
\acmConference[NanoCom '25]{12th ACM International Conference on Nanoscale Computing and Communication}{October 23-25, 2025}{Chengdu, China} 
\acmISBN{978-1-4503-XXXX-X/2018/06}

\usepackage{tikz}
\usetikzlibrary{positioning}
\usetikzlibrary{arrows.meta}
\usetikzlibrary{arrows}
\setcopyright{none}
\begin{document}

\title{Identification for Molecular Communication Based on Diffusion Channel with Poisson Reception Process}

\author{Yaning Zhao}
\affiliation{%
  \institution{TU Braunschweig}
  \city{Braunschweig}
  \country{Germany}
}
\email{yaning.zhao@tu-braunschweig.de}

\author{Luca Miszewski}
\affiliation{%
  \institution{TU Braunschweig}
  \city{Braunschweig}
  \country{Germany}
}
\email{l.miszewski@tu-braunschweig.de}

\author{Christian Deppe}
\affiliation{
  \institution{TU Braunschweig}
  \city{Braunschweig}
  \country{Germany}
}
\email{christian.deppe@tu-braunschweig.de}

\author{Massimiliano Pierobon}
\affiliation{
    \institution{University of Nebraska-Lincoln}
    \city{Lincoln}
    \state{Nebraska}
    \country{USA}
}
\email{maxp@unl.edu}


\begin{abstract}
Molecular communication (MC) enables information exchange at the nano- and microscale, with applications in areas like drug delivery and health monitoring. These event-driven scenarios often require alternatives to traditional transmission.  Identification communication, introduced by Ahlswede and Dueck, offers such an approach, in which the receiver only determines whether a specific message was sent, suiting resource-limited and event-triggered systems. This paper combines MC with identification and proposes a one-dimensional (1D) diffusion-based model. Diffusion noise is modeled as a Poisson process, and a lower bound on channel capacity is derived. Simulations, microscopic, and with short-length deterministic codes, validate theoretical results, including the channel impulse response and error bounds. The findings support the design of practical MC systems, with potential use in testbed development.
\end{abstract}



\maketitle
\renewcommand\footnotetextcopyrightpermission[1]{}
\section{Introduction}

Technological progress has steadily miniaturized devices, enabling the development of nanoscale networks with applications in medi\-cine, biology, and environmental monitoring. These include targeted drug delivery \cite{Allen.2004, Yoo.2011, Muthu.2009, Farokhzad.2009, Singh.2009}, tissue engineering \cite{Griffith.2002, Tayalia.2009}, brain-machine interfaces \cite{Clausen.2009}, and health monitoring \cite{Moritani.2006, Malak.2012, Byrne.2006}. Since electromagnetic communication is often unsuitable for use inside the human body \cite{Farsad.2016, Guo.2015}, alternative methods are needed. One promising approach is molecular communication (MC), which uses chemical signals such as molecules to transmit information, mimicking biological processes such as cell signaling \cite{Nakano.2013}. Realistic and robust system models are essential for designing and evaluating MC systems. These models must accurately capture the complex physical behavior of information carriers. Once established, information-theoretic analysis can quantify system performance \cite{Hsieh.2013, Pierobon.2013}.

While traditional communication theory focuses on decoding transmitted messages, many MC applications are inherently event-driven \cite{cabrera20216g}. In scenarios such as targeted drug delivery, environmental monitoring, or advanced cancer therapies \cite{wilhelm2016analysis}, the receiver's objective is often not to reconstruct the entire message, but rather to ascertain whether a particular event has occurred. This objective aligns with the identification (ID) communication paradigm introduced by Ahlswede and Dueck~\cite{Ahlswede.1989}, wherein the receiver merely verifies whether a specific message was transmitted. In classical discrete memoryless channels (DMCs), ID with randomized encoding enables a doubly exponential growth in the number of distinguishable messages, scaling as $\sim 2^{2^{nR}}$, where $R$ denotes the rate~\cite{Ahlswede.1989}. A similar doubly exponential scaling behavior has also been observed in discrete-time Poisson channels (DTPCs), which are commonly used to model molecular communication (MC) systems~\cite{labidi24}. Deterministic ID (DI), a variant without random encoding, has likewise been studied for DTPCs, demonstrating super-exponential scaling of the form $\sim 2^{n \log(n) R}$~\cite{DIDTPC}. More generally, it has been shown that such super-exponential growth $\sim 2^{n \log(n) R}$ is achievable for memoryless channels with infinite input alphabets and finite output alphabets~\cite{colomer2025deterministic, colomer2025ratereliability}.

 The reception process has been extensively employed in modeling the channel for ID problem. Several alternative channel models have been explored, including binomial channels \cite{salariseddigh2023deterministic}, DTPCs \cite{DIDTPC}, and fading channels \cite{salariseddigh2021deterministic}. Furthermore, randomized ID for DTPCs has also been investigated in \cite{labidi24}.  However, the actual MC process is more comprehensive than the simple DTPC model, highlighting the for a more practical and realistic system representation.
One approach to achieving this is to model the system using a diffusion channel \cite{Pierobon.2013}. In a diffusion-based MC channel, the transmitter releases molecules into the surrounding medium, where they propagate via Brownian motion. In the regime where a large number of molecules are released, commonly referred to as the large-system limit, the molecular distribution in any small volume can be approximated by its expected value. This deterministic behavior is well described by Fick's law of diffusion \cite{Ficklaw}. Information is conveyed to the receiver by modulating the local molecule concentration in its vicinity.

In this paper, we propose a novel MC model that combines a one-dimensional (1D) diffusion channel with a Poisson reception process, which we refer to as the diffusion-based DTPC. This model captures the interaction between diffusion dynamics and stochastic reception events, offering a more realistic representation of communication systems. Furthermore, we derive, for the first time to our knowledge, a lower bound on the DI capacity for this diffusion-based DTPC, advancing both theoretical understanding and providing a foundation for future optimization studies in similar diffusion-driven communication environments.

\textit{Outline}: The remainder of the paper is organized as follows. Section~\ref{sec:main} introduces the system model and presents the main results. Section~\ref{sec:proof} provides the proof of the lower bound on the DI capacity. Section~\ref{sec:sim} presents numerical results that support our theoretical findings. Finally, Section~\ref{sec:conclusion} concludes the paper and discusses potential directions for future research.

\section{System Model and Main Result}
\label{sec:main}
\begin{figure}[H]
    \centering
    \includegraphics[width=0.8\linewidth]{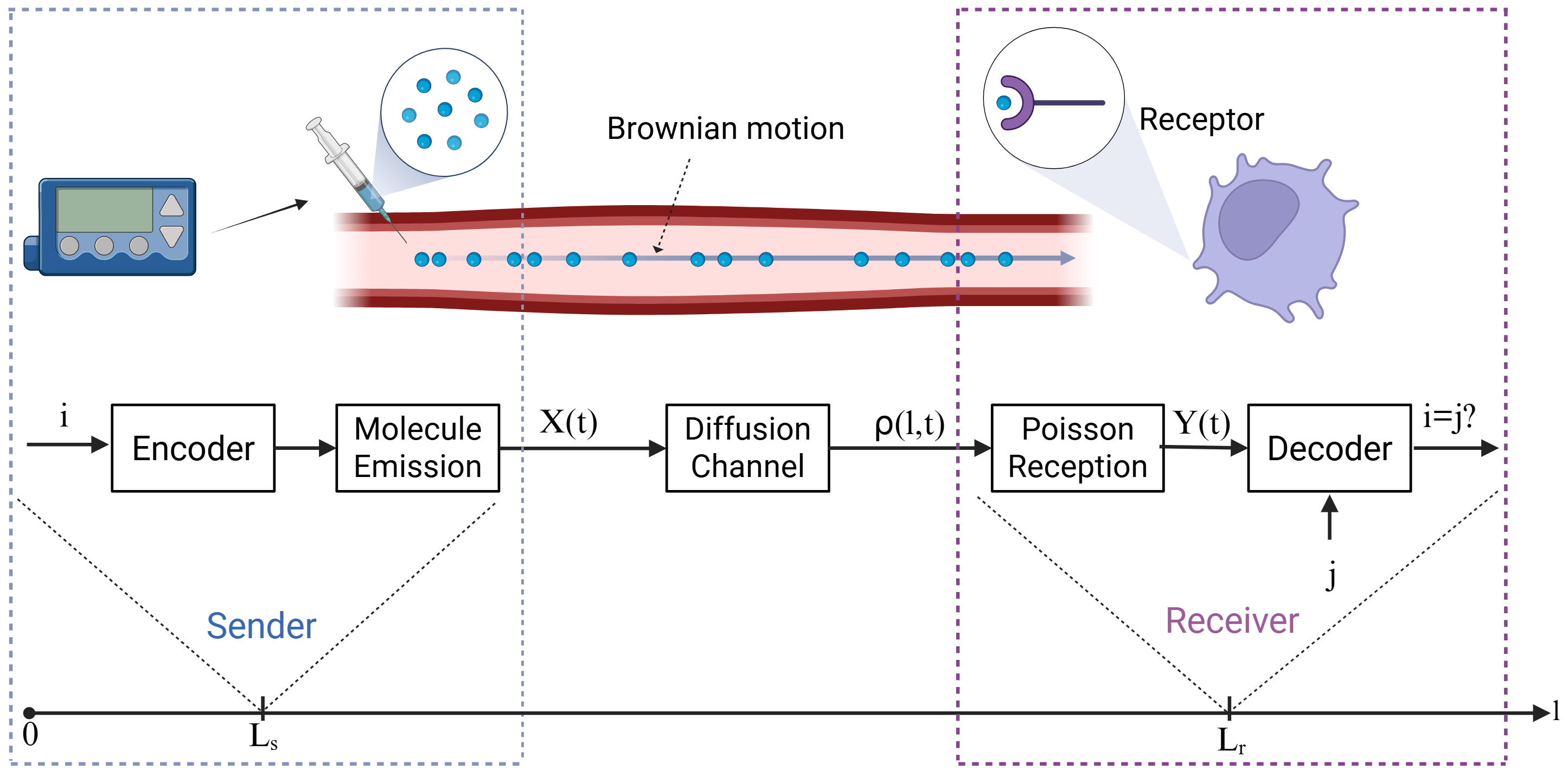}
    \caption{DI over diffusion-based Poisson channel}
    \Description{system model: Di over diffusion-based Poisson channel}
    \label{fig:system_model}
\end{figure}

Consider DI over a diffusion-based DTPC $\mathcal{P}=\left(\mathcal{X},\mathcal{Y},W_{\mathcal{P}}(\cdot|\cdot)\right)$ consisting of an input alphabet $\mathcal{X}\subset\mathbb{R}_0^{+}$, and output alphabet $\mathcal{Y}\subset \mathbb{N}_0$, a Poissonian probability mass function (pmf) $W_{\mathcal{P}}$ on $\mathcal{Y}$, as illustrated in Figure \ref{fig:system_model}. We assume that the sender is located at the origin of the 1D coordinate system, that is, $L_S=0$, and the receiver is located at $L_R$. Let the concentration of the molecule in the medium be indicated by $\rho(l,t)$, where $l$ denotes the coordinate and $t\in[0,\infty)$ is the time. The sender releases $x(t)$ molecules by on-off keying (OOK) in the medium for all $t\in[0,\infty)$. Observe that the sender is subject to a peak power constraint. We assume that the maximum number of released molecules is restricted by $A$, i.e.,
\begin{align}
    0\leq x(t)\leq A, \quad \forall t\in[0,\infty).\nonumber
\end{align}
The diffusion channel is characterized by the following equations:
\begin{align}
    \left\{
    \begin{array}{cc}
         \frac{\partial\rho(l,t)}{\partial t}=D\frac{\partial^2\rho(l,t)}{\partial^2l}& \text{(Fick's second law)} \\
         \rho(l,0)=\delta(l)&\text{(initial condition)} \\
         \rho(\infty,t)=0&\text{(first boundary condition)}\\
         D\frac{\partial\rho(l,t)}{\partial l}\bigg|_{l=L_R}=0&\text{(second boundary condition)}
    \end{array}
    \right.\nonumber,
\end{align}
where $D$ is the diffusion coefficient.
Apply Green's function for 1D boundary-free diffusion equation \cite{Gohari.2016}, we get for all $t>0$,
\begin{align}
    \rho(l,t)=\frac{1}{\sqrt{4\pi Dt}}\exp{\left(-\frac{l^2}{4Dt}\right)}+\frac{1}{\sqrt{4\pi Dt}}\exp{\left(-\frac{(l-2L_R)^2}{4Dt}\right)}.\nonumber
\end{align}
Define the mean number of absorbed molecules as $\lambda_t\triangleq x_t\tilde{\lambda}_t$, where $x_t$ is the number of released molecules and $\tilde{\lambda}_t\in[0,1]$ is the absorbing probability given by
\begin{align}
    \tilde{\lambda}_t
    &=1-\int_{0}^{L_R}{\rho(l,t)}dl\nonumber\\
    &=\text{erfc}\left(\frac{L_R}{\sqrt{4Dt}}\right),\nonumber
\end{align}
where $\text{erfc}(x)=\frac{2}{\sqrt{\pi}}\int_{x}^\infty {e^{-t^2}}dt$ is known as the complementary error function.

For all $t>0$, $x_t\in\mathcal{X}$, and $y_t\in\mathcal{Y}$, the diffusion-based Poisson channel is characterized by
\begin{align}
    W_{\mathcal{P}}\left(y_t|x_t\right)
    &=\frac{\lambda_t^{y_t}e^{-\lambda_t}}{y_t!}\nonumber\\
    &=\frac{\left(\text{erfc}\left(\frac{L_R}{\sqrt{4Dt}}\right)\cdot x_t\right)^{y_t}\text{exp}\left(-\text{erfc}\left(\frac{L_R}{\sqrt{4Dt}}\right)\cdot x_t\right)}{y_t!}.\nonumber
\end{align}
In the following, we define a DI code for diffusion-based DTPC.

\begin{definition} (DI code) 
    Let $\lambda_1,\lambda_2\geq0$, and $\lambda_1+\lambda_2<1$. Then, an $(n,N,\lambda_1,\lambda_2)$ DI code for diffusion-based DTPC $\mathcal{P}$ under maximal molecule release rate $A$ is a family $\left\{\left(\boldsymbol{u}_i,\mathcal{D}_i\right)|i\in[N]\right\}$ of codewords $\boldsymbol{u}_i\in\mathcal{X}^n$ satisfying maximum molecule release rate constraint $0\leq u_{i,t}\leq A$ for all $t\in[n]$ and decoding regions $\mathcal{D}_i\in{\mathbb{N}_0^+}^n$; such that the Type I error and the Type II error satisfy
    \begin{align}
        P_{e,1}(i)&\triangleq W_{\mathcal{P}}^n\left(\mathcal{D}_i^c|\boldsymbol{u}_i\right)\leq \lambda_1,\quad \forall i\in[N]\nonumber\\
        P_{e,2}(i,j)&\triangleq W_{\mathcal{P}}^n\left(\mathcal{D}_j|\boldsymbol{u}_i\right)\leq \lambda_2, \quad \forall i\neq j\in [N].\nonumber
    \end{align}
\end{definition}
Each code can be characterized by the rate, suitably defined for DI in the super-exponential scale as $R= \frac{\log{N}}{n\log{n}}$. A rate is said to be achievable, if there exists an $\left(n,N,\lambda_1,\lambda_2\right)$ DI code that can attain it. Given a diffusion-based DTPC $\mathcal{P}$, we define the super-exponential capacity $C^d_{ID}(\mathcal{P})$ as the supremum of all achievable rate in the limit of infinite channel uses:
\begin{align}\label{eq:capacity}
    C_{ID}^d(\mathcal{P})=\inf_{\lambda_1,\lambda_2>0}\liminf_{n\to\infty}\frac{1}{n\log{n}}\log{N_{ID}^d(\lambda_1,\lambda_2)},
\end{align}
where $N_{ID}^d(\lambda_1,\lambda_2)$ is the maximum number of words in the code.

In this work we prove the following:

\begin{theorem}\label{thm:main}
    The DI capacity of a diffusion-based DTPC $\mathcal{P}$ under maximum molecule release rate constraint $0\leq x_t\leq A$ is upper-bounded by
    \begin{align}
         C_{ID}^d(\mathcal{P})\geq \frac{1}{4}.\nonumber
    \end{align}
\end{theorem}

\section{Proof of Theorem \ref{thm:main}}
\label{sec:proof}
\begin{figure}[H]
    \centering
    \begin{tikzpicture}[scale=1.3]

%


\node at (3.5,1.5) {\(A\)};
\node at (1.5,-0.5) {\(A\)};

\node at (-0.25,-0.25) {$r_0$};

\foreach \x/\y in {0.7/0.7, 1.5/0.7, 2.3/0.7,
                   0.7/1.5, 1.5/1.5, 2.3/1.5,
                   0.7/2.3, 1.5/2.3, 2.3/2.3} {
    \draw[thick, fill=blue!10] (\x,\y) circle(0.4);
    \fill[black] (\x,\y) circle(0.015); 
}

\foreach \x/\y in {0.13/0.13, 2.87/0.13, 0.13/2.87, 2.87/2.87, 0.005/1.1, 0.005/1.9, 1.1/0.005, 1.9/0.005, 1.1/2.995, 1.9/2.995, 2.995/1.1, 2.995/1.9 } {
    \filldraw[fill=blue!10, draw=black] (\x,\y) circle(0.4);
    \fill[black] (\x,\y) circle(0.015); 
}

Square box
\draw[line width=1.2pt] (0,0) rectangle (3,3);

\draw[thick] (0.13,0.13) -- (-0.15,-0.15);

\end{tikzpicture}
    \caption{Sphere packing technology for achievability proof}
    \Description{Visualization of a packing of hyperspheres with radius $r_0$ inside a hypercube with edge length A. Each codeword lie in the center of the hypersphere}
    \label{fig:packing}
\end{figure}
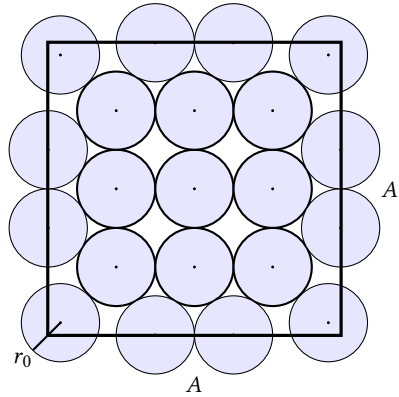

We present the proof by extending the sphere packing technique used for the achievability proof in \cite{DIDTPC} to our system model. Let $\mathscr{P}$ denotes a sphere packing, i.e., an arrangement of non-overlapping spheres $S_{\boldsymbol{u}_i}\left(n,r_0\right)$, $i\in[N]$, that are packed inside the large cube $Q_0(n,A)$ with edge length $A$, as illustrated in Fig. \ref{fig:packing}. 
The radius $r_0$ is of hyperspheres given by: $r_0\triangleq \sqrt{a}n^{\frac{1+b}{4}}$, where $a$ and $b$ are constants with $a>0$ and $b\in[0,1]$. The number of codewords $N$ is given by
\begin{align}
    N
    &=\frac{\text{Vol}\left[\bigcup_{i=1}^{N}S_{\boldsymbol{u}_i}(n,r_0)\right]}{\text{Vol}\left[S_{\boldsymbol{u}_i}(n,r_0)\right]}\nonumber\\
    &=\frac{\Delta_n(\mathscr{P})\text{Vol}[Q_0(n,A)]}{\text{Vol}[S_{\boldsymbol{u}_i}(n,r_0)]}\nonumber,
\end{align}
where $\Delta_n(\mathscr{P})\triangleq \frac{\text{Vol}\left[\bigcup_{i=1}^{N}S_{\boldsymbol{u}_i}\left(n,r_0\right)\right]}{\text{Vol}\left[Q_0(n,A)\right]}$ denotes the packing density. We introduce the following lemma:
\begin{lemma}\cite{cohn2010order} 
    Every saturated pacing has a density $\Delta_n(\mathscr{P})\geq 2^{-n}$.
\end{lemma}
The number of hyperspheres can be lower-bounded by
\begin{align}
    N
    & \geq \frac{2^{-n}\text{Vol}\left[Q_0(n,A)\right]}{\text{Vol}\left[S_{\boldsymbol{u}_i}(n,r_0)\right]} \nonumber\\
    &\overset{(a)}{\geq} \frac{2^{-n}A^n\Gamma\left(\frac{n}{2}+1\right)}{\sqrt{\pi^n}r_0^n}\nonumber\\
    &\geq \frac{2^{-n}A^n(\lfloor\frac{n}{2}\rfloor\log{\lfloor\frac{n}{2}\rfloor}-\lfloor\frac{n}{2}\rfloor\log{e}+o(n))}{\sqrt{\pi^n}r_0^n},\nonumber
\end{align}
where $(a)$ follows from the volumes $\text{Vol}\left[Q_0(n,A)\right]=A^n$ and $\text{Vol}\left[S_{\boldsymbol{u}_i}\left(n,r_0\right)\right]=\frac{\sqrt{\pi^n}r_0^n}{\Gamma\left(\frac{n}{2}+1\right)}$, where $\Gamma(\cdot)$ denotes the Gamma function defined by $\Gamma(\frac{n}{2}+1)=\frac{n}{2}!$.
Take the logarithm
\begin{align}
    \log{N} \geq n\log{n}\left[{\frac{1}{4}(1-b)+o(n)}\right].\nonumber
\end{align}
Choose $b\to 0$ and insert this into Eq. \eqref{eq:capacity} we obtain the claimed capacity lower bound.

For code construction, a distance decoder is used and for $i\in[N]$, the decoding region is given by
\begin{align}
    \mathcal{D}_i\triangleq \left\{y^n\in\mathcal{Y}^n: \big|d(y^n,\boldsymbol{u}_i)\big|\leq \delta_n\right\},\nonumber
\end{align}
where $d(y^n,\boldsymbol{u}_i)=\frac{1}{n}\sum_{t=1}^n\left[(y_t-\lambda_t)^2-y_t\right]$ and $\delta_n\triangleq ac\left(\min_{t}\tilde{\lambda}_t\right)^2n^{\frac{b-1}{2}}$ with constant $c\in(0,2)$.

For all $i\in[N]$, the type I error can be upper-bounded by
\begin{align}
    P_{e,1}(i)
    &=W^n_\mathcal{P}(\mathcal{D}_i^c\big|\boldsymbol{u}_i)\nonumber\\
    &= \text{Pr}\left[|{d(Y^n,\boldsymbol{u}_i)|>\delta_n}\big|\boldsymbol{u}_i\right]\nonumber\\
    &=\text{Pr}\left[\bigg|\frac{1}{n}\sum_{t=1}^n\left[\left(Y_t-\tilde{\lambda}_tu_{i,t}\right)^2-Y_t\right]\bigg|>\delta_n\right]\nonumber\\
    &\overset{(a)}{=}\text{Pr}\left[\left|\frac{1}{n}\sum_{t=1}^n\left[\left(Y_t- \tilde{\lambda}_tu_{i,t}\right)^2-Y_t\right]\right.\right.\nonumber\\
    &\left.\left.\quad -\mathbb{E}\left\{\frac{1}{n}\sum_{t=1}^n\left[\left(Y_t- \tilde{\lambda}_tu_{i,t}\right)^2-Y_t\right]\right\}\right|>\delta_n \right]\nonumber\\
    &\overset{(b)}{\leq} \frac{\text{Var}\left\{\frac{1}{n}\sum_{t=1}^n\left[\left(Y_t- \tilde{\lambda}_tu_{i,t}\right)^2-Y_t\right]\right\}}{\delta_n^2}\nonumber,
\end{align}
where $(a)$ follows from 
\begin{align}
    \mathbb{E}\left[\frac{1}{n}\sum_{t=1}^n\left[\left(Y_t- \tilde{\lambda}_tu_{i,t}\right)^2-Y_t\right]\right]
    &=\frac{1}{n}\sum_{t=1}^n\mathbb{E}\left\{\left(Y_t-\mathbb{E}[Y_t]\right)^2-\tilde{\lambda}_tu_{i,t}\right\}\nonumber\\
    &=\frac{1}{n}\sum_{t=1}^n\left\{\text{Var}\left[Y_t\right]-\tilde{\lambda}_tu_{i,t}\right\}\nonumber\\
    &=0\nonumber,
\end{align}
and $(b)$ follows from the Chebyshev's inequality. Thus, the Type I error can be bounded by
\begin{align}
    P_{e,1}(i)
    &\leq\frac{\text{Var}\left\{\frac{1}{n}\sum_{t=1}^n\left[\left(Y_t-\tilde{\lambda}_tu_{i,t}\right)^2-Y_t\right]\right\}}{\delta_n^2}\nonumber\\
    &\leq\frac{\frac{1}{n^2}\sum_{t=1}^n\left\{\mathbb{E}\left[(Y_t-\tilde{\lambda}_tu_{i,t})^4\right]-\tilde{\lambda}_tu_{i,t}\right\}}{\delta_n^2}\nonumber\\
    &=\frac{\frac{1}{n^2}\sum_{t=1}^n\left(\mu_{4}(u_{i,t})-\tilde{\lambda}_tu_{i,t}\right)}{\delta_n^2}\nonumber\\
    &\leq \frac{3A}{n^2\delta_n^2}\sum_{t=1}^n\tilde{\lambda}_t^2\nonumber\\
    &\leq \frac{3A}{a^2c^2(\min_{t}\tilde{\lambda}_t)^4n^b} = o(n)\nonumber,
\end{align}
where $\mu_4(u_{i,t})=\mathbb{E}\left[\left(Y_t-\mathbb{E}\left[Y_t\right]\right)^4\right]=\tilde{\lambda}_tu_{i,t}+3\lambda^2(u_{i,t})$ denotes the fourth central moment of $Y_t$. We can achieve arbitrary small Type I error probability with sufficient large $n$.

For all $i\neq j\in[N]$, the type II error can be upper-bounded by
\begin{align}
    P_{e,2}(i,j)
    &=W^n_{\mathcal{P}}(\mathcal{D}_j|\boldsymbol{u}_i)\nonumber\\
    &=\text{Pr}\left[\left|d(Y^n,\boldsymbol{u}_j)\right|\leq \delta_n\big|\boldsymbol{u}_i\right]\nonumber\\
    &=\text{Pr}\left[\Bigg|\frac{1}{n}\sum_{t=1}^n\left[\left(Y_t-\lambda_t(u_{j,t})\right)^2-Y_t\right]\Bigg|\leq \delta_n\bigg| \boldsymbol{u}_{i}\right]\nonumber\\
    &\leq\text{Pr}\left[\frac{1}{n}\sum_{t=1}^n\left[\left(Y_t-\tilde{\lambda}_tu_{i,t}\right)^2+\tilde{\lambda}^2_t\left(u_{i,t}-u_{j,t}\right)^2\right]\right.\nonumber\\
    &\quad +\frac{2}{n}\sum_{t=1}^n\tilde{\lambda}_t\left(Y_t-\tilde{\lambda}_tu_{i,t}\right)\left(u_{i,t}-u_{j,t}\right)-\frac{1}{n}\sum_{t=1}^n Y_t 
    \left.\leq \delta_n\bigg|\boldsymbol{u}_{i}\right]\nonumber
    \end{align}
    \begin{align}
    &\overset{(a)}{\leq} \text{Pr}\left[\left\{\bigg|\frac{2}{n}\sum_{t=1}^n\tilde{\lambda_t}(Y_t-\tilde{\lambda_t}u_{i,t})(u_{i,t}-u_{j,t})\bigg|> \delta_n\right\}\right.\nonumber\\
    &\quad\left.\bigcap\left\{\frac{1}{n}\sum_{t=1}^n\left[(Y_t-\tilde{\lambda}_tu_{i,t})^2+\tilde{\lambda}_t^2(u_{i,t}-u_{j,t})^2-Y_t\right]\leq 0\right\} \bigg| \boldsymbol{u}_{i}\right]\nonumber\\
    &\quad + \text{Pr}\left[\left\{\bigg|\frac{2}{n}\sum_{t=1}^n\tilde{\lambda_t}(Y_t-\tilde{\lambda_t}u_{i,t})(u_{i,t}-u_{j,t})\bigg|\leq \delta_n\right\}\right.\nonumber\\
    &\quad \bigcap\left.\left\{\frac{1}{n}\sum_{t=1}^n\left[(Y_t-\tilde{\lambda}_tu_{i,t})^2+\tilde{\lambda}_t^2(u_{i,t}-u_{j,t})^2-Y_t\right\}\right]\leq 2\delta_n\bigg|\boldsymbol{u}_{i}\right]\nonumber\\
    &\leq \text{Pr}\left[\bigg|\frac{2}{n}\sum_{t=1}^n\tilde{\lambda_t}(Y_t-\tilde{\lambda_t}u_{i,t})(u_{i,t}-u_{j,t})\bigg|> \delta_n \bigg| \boldsymbol{u}_{i}\right]\nonumber\\
    &\quad + \text{Pr}\left[\frac{1}{n}\sum_{t=1}^n\left[(Y_t-\tilde{\lambda}_tu_{i,t})^2+\tilde{\lambda}_t^2(u_{i,t}-u_{j,t})^2-Y_t\right]\leq 2\delta_n\bigg|\boldsymbol{u}_{i}\right]\nonumber\\
    &\overset{(b)}{\leq}\frac{\text{Var}\left[\sum_{t=1}^n\frac{2}{n}\tilde{\lambda}_t(Y_t-\tilde{\lambda}_tu_{i,t})(u_{i,t}-u_{j,t})\bigg|\boldsymbol{u}_{i}\right]}{\delta_n^2}\nonumber\\
    &\quad +\text{Pr}\left\{\frac{1}{n}\sum_{t=1}^n\left[\left({Y_t-u_{i,t}}\right)^2-Y_t\right]+\frac{1}{n}{\left(\min_{t}{\tilde{\lambda}_t}\right)^2||\boldsymbol{u}_{i}-\boldsymbol{u}_{j}}||^2\leq 2\delta_n\bigg|\boldsymbol{u}_{i}\right\}\nonumber\\
    &\leq \frac{4\left(\max_{t}\tilde{\lambda}_t\right)^2}{n^2\delta_n^2}\sum_{t=1}^n\left(u_{i,t}-u_{j,t}\right)^2 \text{Var}\left[Y_t-\tilde{\lambda_t}u_{i,t}\big|u_{i,t}\right]\nonumber\\
    &\quad + \text{Pr}\left[\bigg|\frac{1}{n}\sum_{t=1}^n\left[Y_t-\left(Y_t-u_{i,t}\right)^2\right]\bigg|\geq \frac{1}{n}\left(\min_{t}\tilde{\lambda}_t\right)^2\left(2r_0\right)^2-2\delta_n\bigg|\boldsymbol{u}_{i}\right]\nonumber\\
    &\overset{(c)}{\leq} \frac{8A^3\left(\max_{t}\tilde{\lambda}_t\right)^3}{n\delta_n^2}+\frac{3A}{n\left[\frac{4}{n}\left(\min_t\tilde{\lambda}_t\right)^2r_0^2-2\delta_n\right]^2}\nonumber\\
    &\leq \frac{8A^3\left(\max_t{\tilde{\lambda}_t}\right)^3}{ac\left(\min_t{\tilde{\lambda}}_t\right)^2n^{b+1}}+\frac{3A}{4a^2\left(\min_t{\tilde{\lambda}_t}\right)^4n^b\left(2-cn^{\frac{b-1}{2}}\right)}= o(n),\nonumber
\end{align}
where $(a)$ follows from low of total probability, $(b)$ and $(c)$ follows from Chebyshev's inequity. We can achieve arbitrary small Type II error with sufficient large code length $n$. This completes the proof of Theorem \ref{thm:main}.

\section{Numerical Result}
\label{sec:sim}
This section presents simulation results for molecular diffusion modeled by macroscopic methods and the performance evaluation of a DI coding scheme.
\subsection{Macroscopic Simulation}

With a large number of molecules, diffusion in the channel is modeled by Fick's second law. The simulation uses the finite-element method with spatial and time grids. In 1D, the grid is a series of discrete points. The molecular concentration $\rho(l,t)$ is updated iteratively as follows:
\begin{align}
    \rho(l, t+\Delta t) = \rho(l,t) + \frac{D \Delta t}{(\Delta l)^2}  \bigg( \rho(l-\Delta l,t) -2\rho(l, t) +\rho(l+\Delta l, t)\bigg), \nonumber
\end{align}
where $\Delta l$ and $\Delta t$ represent the spatial and time resolution, respectively, and D is the diffusion coefficient. 
The factor $\frac{D \Delta t}{(\Delta l)^2}$ is the model diffusion coefficient and must be below 0.5 for stability. Higher resolution improves accuracy but increases computation. The simulation initializes arrays for concentration, absorption, and absorption rate. At $t=0s$, all molecules are released at one point. Each timestep updates concentrations via diffusion, removing molecules at the receiver to model absorption. Parameters are in Table \ref{tab:parametersmacsim}.

\begin{figure}[H]
    \centering
    \includegraphics[width=0.65\linewidth]{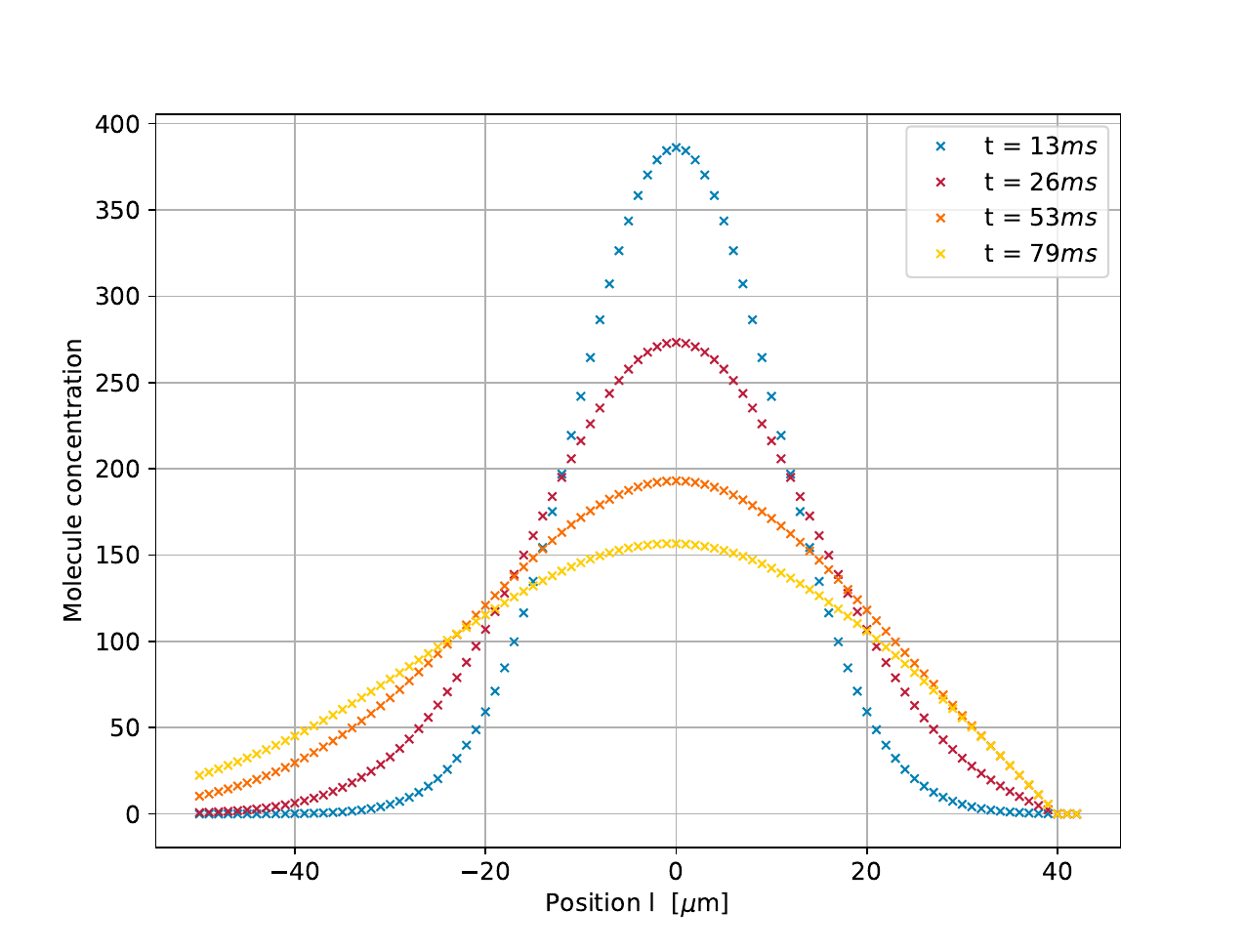}
    \caption{Molecule concentration for different times}
    \Description{}
    \label{fig:mac_sim_concentration}
\end{figure}

\begin{figure}[H]
    \centering
    \includegraphics[width=0.6\linewidth]{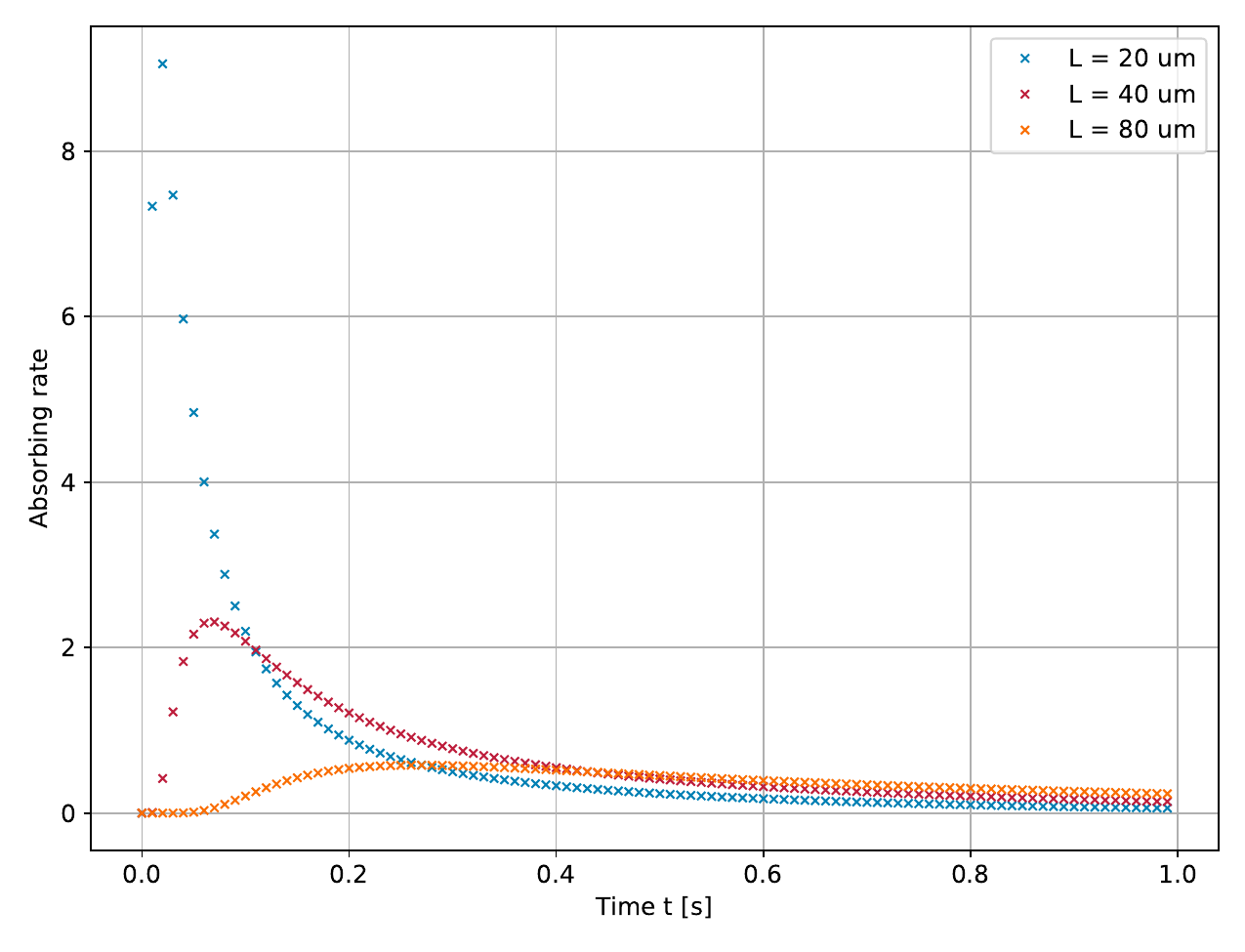}
    \caption{Absorbing rate for different receiver positions}
    \Description{}
    \label{fig:mac_sim_rate}
\end{figure}

\begin{figure}[H]
    \centering
    \includegraphics[width=0.65\linewidth]{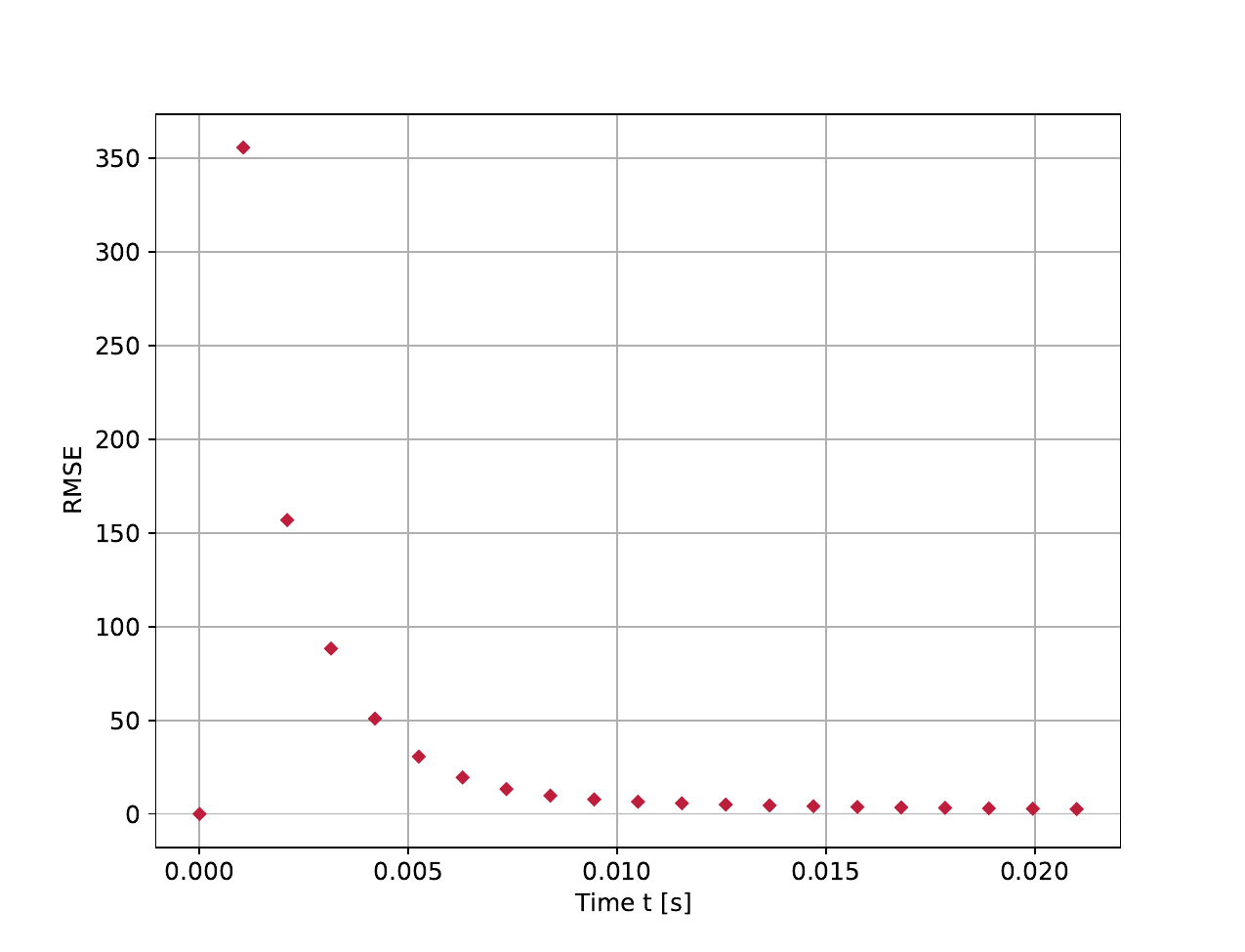}
    \caption{RMSE between macroscopic simulation and numerical solution}
    \Description{}
    \label{fig:mac_sim_rmse}
\end{figure}

\begin{table}[H]
    \caption{Parameter values for the macroscopic simulation.}
    \centering
    \begin{tabular}{|c|l|c|}
        \hline
        \textbf{Parameter} & \textbf{Name} & \textbf{Value} \\
        \hline
         $x_t$& Number of molecules & $10,000$ \\        
        $D$ & Diffusion coefficient & $4 \times 10^{-9}\frac{m^2}{s}$ \\
        $L_S$  & Sender position & 0m \\
        $L_R$  & Receiver position & $40 \times 10^{-6}m$ \\
        $\Delta t$ & Time grid resolution & $1 \times 10^{-4}s$ \\
        $\Delta l$ & Space grid resolution & $1 \times 10^{-6}m$ \\
        \hline
    \end{tabular}
    \label{tab:parametersmacsim}
\end{table}
A large number of molecules enables concentration-based modeling using standard diffusion and receiver distance values. Spatial and temporal resolutions are set as fine as possible for accuracy and stability. Figure \ref{fig:mac_sim_concentration} shows the concentration over time. At $t=13ms$, absorption is minimal, and the profile matches the boundary-free Green's function. Over time, diffusion flattens the profile, and absorption shifts the peak from the origin.
Figures \ref{fig:mac_sim_rate} show the absorption rate, calculated as the derivative of the number of absorbed molecules. Receiver positions from $20\mu m$ to $80\mu m$ are simulated. The absorption curves follow the complementary error function from Section~\ref{sec:main}. Greater sender-receiver distance delays molecule arrival and flattens the absorption profile, reflected in a lower, later peak in the absorption rate. The time of the maximum rate is given by $\hat{t} = \frac{L^2}{6D}$.
The root mean squared error (RMSE) is used to compare the simulated concentration profile with that from Section~\ref{sec:main}. Figure \ref{fig:mac_sim_rmse} shows the RMSE over time. At $t=0s$, the RMSE is zero since both start with molecules released at the sender. After the first step, RMSE peaks, then rapidly decreases toward zero as the simulation converges to the numerical concentration profile.

\subsection{Identification Code}
\begin{figure}[H]
    \centering 
    \includegraphics[width=0.7\linewidth]{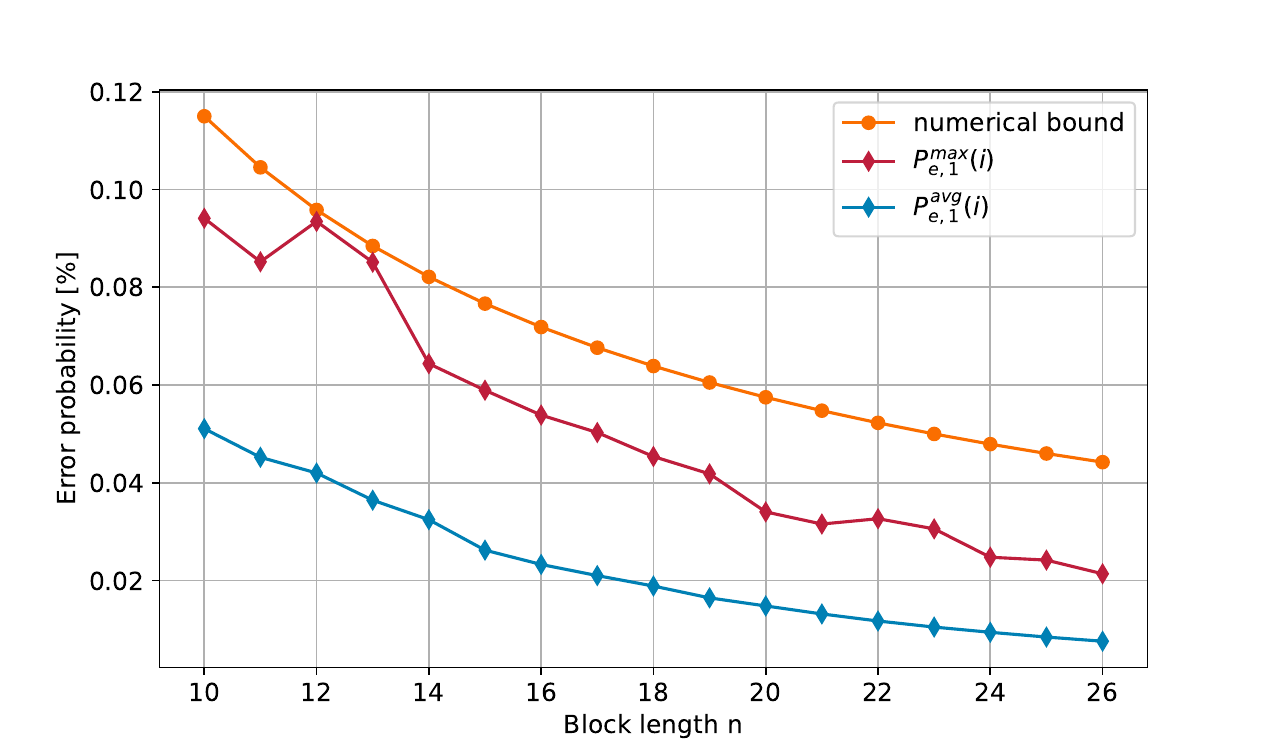}
    \caption{Average and maximum error probability for the  Type I and Type II error and numerical bound versus block length $n$}
    \Description{}
    \label{fig:di_error_1}
\end{figure}

\begin{figure}[H]
    \centering
    \includegraphics[width=0.7\linewidth]{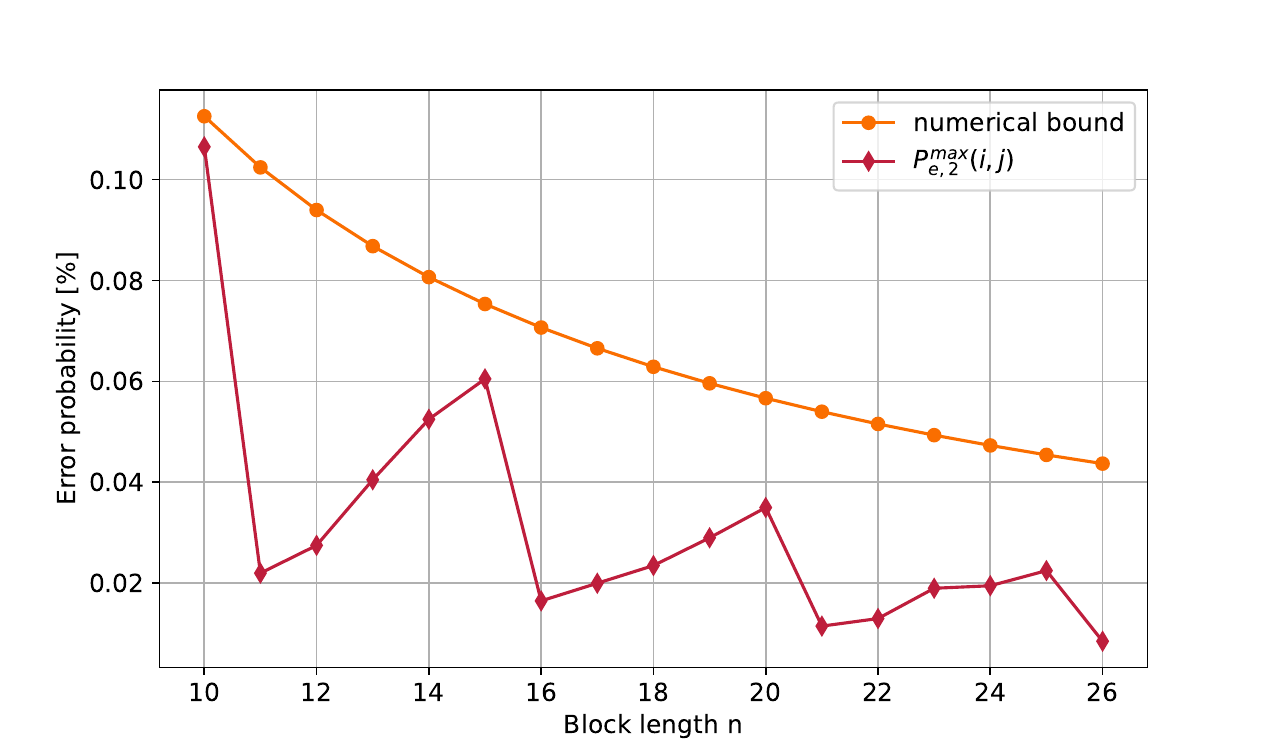}
    \caption{Maximum error probability for the Type II error and numerical bound versus block length $n$}
    \Description{}
    \label{fig:di_error_2}
\end{figure}

\begin{figure}[H]
    \centering
    \includegraphics[width=0.7\linewidth]{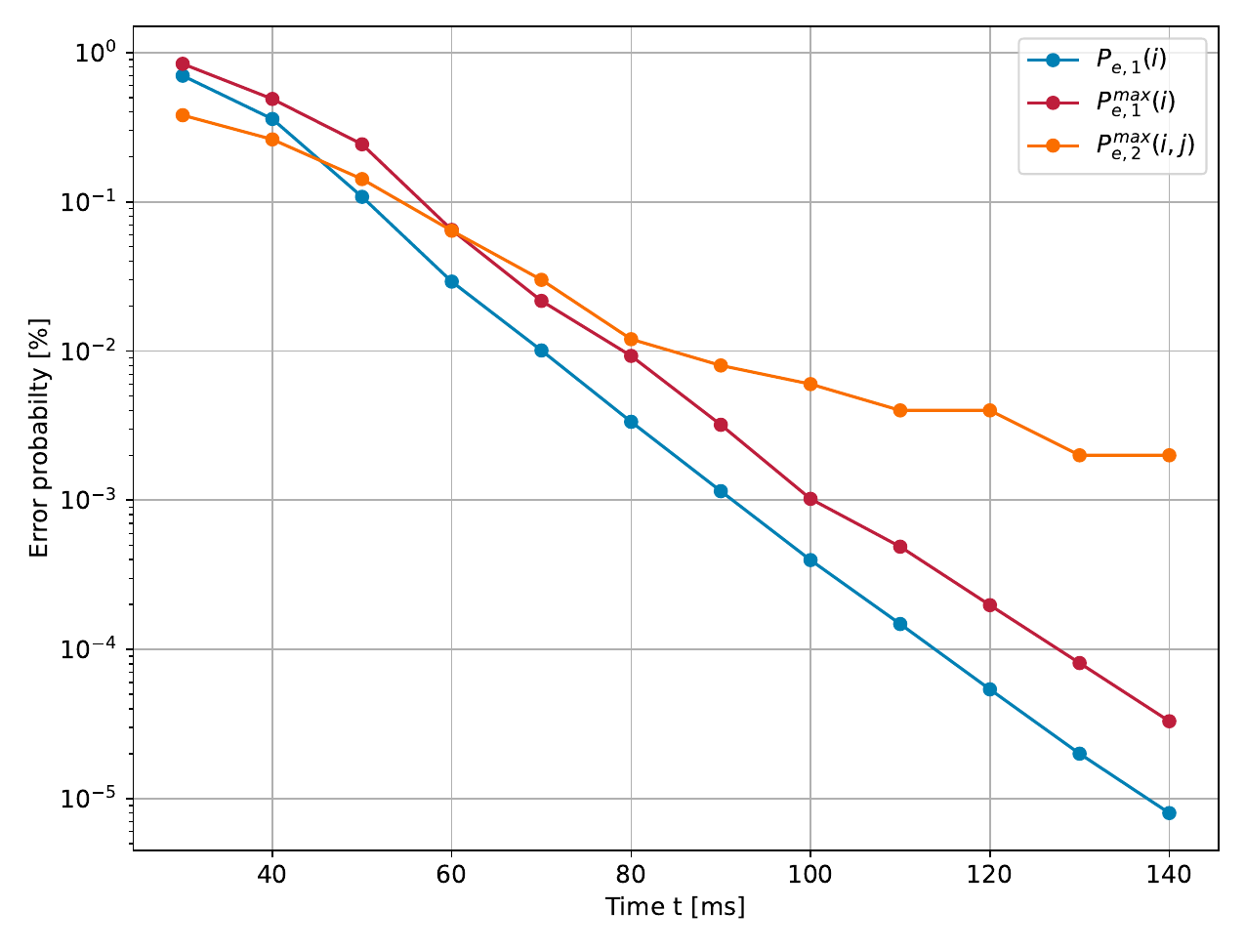}
    \caption{Error probabilities to progressed diffusion time}
    \Description{}
    \label{fig:err_vs_t}
\end{figure}

The DI code is built using a randomized binary codeword based on OOK. The number of codewords $N$ is given by $N=2^{n\log{(n)}R}$. Each codeword must be at least $2r_0$ away from all previously accepted codewords. After constructing the code book, Type I and II error probabilities are evaluated, as illustrated in Fig. \ref{fig:di_error_1} and \ref{fig:di_error_2}. For Type I error, each codeword is transmitted via Monte Carlo simulation, and the decoder checks if the received signal matches its decoding set. The average and maximum Type I errors are then computed. Type II error is assessed by checking whether the signal incorrectly matches decoding sets of other codewords ($i\neq j$). All parameters are listed in Table \ref{tab:parameters_di}.
\begin{table}[H]
    \caption{Parameter values for deterministic identification code}
    \centering   
    \begin{tabular}{|c|l|c|}
        \hline
        \textbf{Parameter} & \textbf{Name}  & \textbf{Value}\\
        \hline

        $x_t$ & Number of molecules & 100\\
         $\tilde{\lambda}_t$ & Absorbing probability & 0.083\\
        $n$ & block length & 10-26 \\
        $R$ & Code rate & 0.1\\
        $a$ & Codebook construction parameter & 500 \\
        $b$ & Codebook construction parameter & 0.99\\
        $c$ & Decoding parameter & 1.5 \\
        $iter_1$ & Iteration for Type I error & 100,000 \\
        $iter_2$ & Iterations for Type II error & 2,000  \\

        \hline
    \end{tabular}
\label{tab:parameters_di}
\end{table}
The maximum Type I and II error probabilities decrease with block length, slightly outperforming the numerical bound. The sawtooth pattern in the Type II error stems from the OOK scheme: codewords lie on the hypercube edges, and the hypersphere radius $r_0$ increases with $n$. Every fifth increase in $n$ causes 
$r_0$ to cross an additional edge, allowing greater Hamming separation. 
Previous simulations used a fixed absorption probability based on the time $\hat{t}$ when the absorption rate peaks. However, this probability actually equals $\text{erfc}\left(\frac{L_R}{\sqrt{4Dt}}\right)$. To examine its impact, a simulation with fixed block length and varying diffusion time was run. The resulting Type I and II error trends are shown in Figure \ref{fig:err_vs_t}. Both error probabilities show a decrease with diffusion time. The maximum slope of Type I error probability is achieved after $t=60ms$ and is nearly constant after this time. Whereas you can see an exact maximum slope for the Type II error probability at around $t=70ms$. This is close to the time, where the maximum absorbing rate is achieved at $\hat{t}=66.6ms$.


\section{Discussion and Conclusion}
\label{sec:conclusion}
This work bridges molecular communication (MC) and identification communication to address the needs of event-driven, resource-constrained nano- and microscale systems. By modeling 1D diffusion noise and Poisson process, we derive a theoretical lower bound on channel capacity and validate it through microscopic simulations and short-length deterministic codes. The results, including the channel impulse response and error performance, confirm the feasibility of identification-based MC and lay a foundation for the development of efficient, practical communication schemes in biomedical and nanonetwork applications. Future research directions include addressing inter-symbol interference (ISI), establishing converse results for capacity, and designing constructive discrete identification codes for simulation.

\begin{acks}
C. Deppe, Luca Miszewski, and Y. Zhao acknowledge financial support from the Federal Ministry of Education and Research of Germany (BMBF) through the program “Souverän. Digital. Vernetzt.” as part of the joint project 6G-life (project identification numbers: 16KISK263). C. Deppe also received partial support under NewCom through Grant 16KIS1005. Furthermore, C. Deppe and Y. Zhao were supported by the DFG within the project DE1915/2-1.
\end{acks}

\bibliographystyle{ACM-Reference-Format}
\bibliography{main}

\appendix

\end{document}